\documentclass[aps,prb,amsfonts,amssymb,epsfig,twocolumn,groupedaddress,showpacs]{revtex4} 
\usepackage{latexsym}
\usepackage[dvips]{graphicx}
\usepackage{epsfig}
\usepackage{amsmath}

\providecommand{\be}{\begin{eqnarray}}
\providecommand{\enn}{\end{eqnarray}}

\begin{document}

\title{Electronic Hong-Ou-Mandel interferometer for multi-mode 
	entanglement detection}

\author{Vittorio Giovannetti}
\affiliation{NEST-CNR-INFM and Scuola Normale Superiore, I-56126 Pisa, Italy}

\author{Diego Frustaglia}
\affiliation{NEST-CNR-INFM and Scuola Normale Superiore, I-56126 Pisa, Italy}

\author{Fabio Taddei}
\affiliation{NEST-CNR-INFM and Scuola Normale Superiore, I-56126 Pisa, Italy}

\author{Rosario Fazio}
\affiliation{International School for Advanced Studies (SISSA),
        via  Beirut 2-4,  I-34014 Trieste, Italy\\
	NEST-CNR-INFM and Scuola Normale Superiore, I-56126 Pisa, Italy}

\date{\today}
\begin{abstract}
We show that multi-mode entanglement of electrons in a mesoscopic conductor can be 
detected by a measurement of the zero-frequency current correlations in an 
electronic Hong-Ou-Mandel interferometer. 
By this means, one can further establish a lower bound to the entanglement of formation of 
two-electron input states.
Our results extend the work of Burkard and Loss [Phys. Rev. Lett. 
{\bf 91}, 087903 (2003)] to many channels and provide a way to test the existence of 
entangled states involving {\em both} orbital and spin degrees of freedom.

\end{abstract}

\maketitle
\section{Introduction}
Differently  from quantum optics, where entangled photons are routinely 
detected by coincidence measurements, the issue of revealing the presence of entangled states in multi-terminal mesoscopic conductors (see, e.g., Ref.~\onlinecite{reviews} and 
references therein) is still an experimental challenge. A number of theoretical 
proposal have been considered so far. Bell-like tests by means of measurements of
current-fluctuations have been analyzed by several authors~\cite{bell_in}. 
Quantum state tomography has been discussed in Ref.~\onlinecite{tom}.
Alternatively, Burkard {\em et al.}~\cite{BLS00}  
suggested that singlet and triplet electron pairs could give rise to 
deviations in the zero-frequency shot noise at the output ports of a 50/50 electronic 
beam splitter (BS) as compared to the value observed for an incoming beam of  
independent electrons. Following this work, the full counting statistics for entangled 
electrons in a BS has been studied in Ref.~\onlinecite{taddei02}. 
The effect of a Rashba spin-orbit term or a rotating magnetic field 
in one of the incoming ports was discussed in Refs.~\onlinecite{egues02-05} 
and~\onlinecite{zhao05}, respectively, and in Ref.~\onlinecite{sanjose06} dephasing through additional reservoirs was included.
While these works considered only singlet or triplet incoming states, in Ref.~\onlinecite{SSB04} the case of states generated in an Andreev double dot entangler at the input ports of the BS has been discussed.

\begin{figure}[t]
\begin{center}
\includegraphics[scale=0.3]{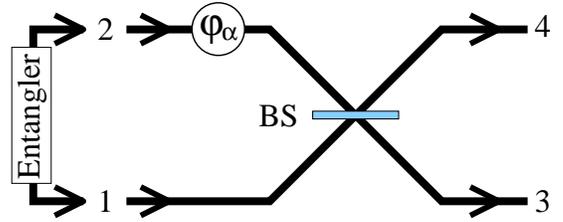}
\end{center}
\caption{Schematics of the apparatus.
        For any energy $E$, 
	two electrons enter from an entangler at ports $1$ and $2$.
	Each port consist of many channels, labeled by a spin ($s$) 
	and orbital ($\ell$) index $\alpha \equiv \{ s,\ell\}$. 
	The BS (with transmissivity $T$ that could be controlled by a gate 
        voltage) mixes the incoming state, leading
	to an outgoing state propagating along ports $3$ and $4$.
	Currents are measured at $3$ and $4$ for different values of 
	a channel-dependent phase shift $\varphi_\alpha$ 
	introduced in port $2$. The presence of spin as well as
	orbital entanglement in the input state (\ref{statoin}) 
	is detected through the Fano factor~(\ref{cc}) for current-noise
	correlations.}
\label{f:figu0}
\end{figure}

Our work is motivated by a recent paper by Burkard and Loss~\cite{burkardloss} where 
the authors derived a lower bound for the entanglement of formation~\cite{BENNETT}
of arbitrary mixed spin states, through shot-noise measurements. In this paper we 
generalize their results to multi-mode input states by introducing the electronic 
analog of the Hong-Ou-Mandel (HOM)~\cite{HOM} optical interferometer.
We assume that a two-electron (possibly entangled) input state is 
injected into a BS whose arms support many propagating  modes, 
including both orbital and spin degrees of freedom.
In this context, we first show that
{\em  zero-frequency} noise measurements allows one to detect entanglement for two-electron input states. 
Then, in the spirit of Ref.~\onlinecite{burkardloss}, we derive a lower bound 
for the entanglement of formation in the case of arbitrary number of incoming channels.
This is relevant from an experimental point of view since it shows 
the possibility of characterizing entanglement in complex scenarios 
which go beyond the paragdimatic two-mode implementation of Burkard and 
Loss original scheme. Moreover, from a theoretical point of view,  
our analysis establishes a clear connection between the detection 
scheme of Ref.~\onlinecite{burkardloss}  and the HOM interferometer 
scheme proposed in optics.

The main ingredient in the present approach is the study of the oscillations of the 
zero-frequency shot noise at the outgoing ports of a BS as 
a function of a controllable phase shift present in one of the incoming ports. 
As we shall see, this technique allows us to detect  spin/orbital as well as more 
complex entangled input states. In view of its simple implementation we believe this 
scheme  can offer a very convenient way for entanglement detection in mesoscopic 
multi-terminal conductors. 

The paper is organized as follows. In Sec.~\ref{sec:interf} we introduce the electronic 
analog of the HOM interferometer and, in Sec.~\ref{sec:witness}, we demonstrate that 
it detects entanglement for two-electron input states.
In Sec.~\ref{sec:lb} we discuss how the zero-frequency  noise measurements
can be related to the entanglement of formation of the input 
state.  The paper finally ends with conclusions and remarks in
Sec.~\ref{sec:conclusion}.

\section{The HOM interferometer}~\label{sec:interf}

In the HOM experiment~\cite{HOM} pairs of multi-mode photons propagating along two 
distinct optical paths interfere at a 50/50 BS after being phase shifted
through  controllable delays. The presence of quantum entanglement in the 
input state of the photons can be recovered by studying the coincidence counts at the 
output ports of the BS as a function of the relative delay
between optical paths. In the electronic analog of the HOM  interferometer (e-HOM) 
metallic conductors play the role of photonic paths and  current correlations 
that of coincidence counts. Apart from the expected differences related to the statistics,
the e-HOM interferometer maintains the same entanglement-detection
capability of its optical counterpart.

The e-HOM is sketched in Fig.~\ref{f:figu0}. 
Pairs of electrons of a given energy $E$ above the Fermi sea, prepared in 
a (possibly entangled) initial state, enter the interferometer 
from the input ports $1$ and $2$. Electrons passing 
through the port $2$ undergo an additional, controllable, phase shift (white 
circle in Fig.~\ref{f:figu0}) before impinging on the BS. 
Zero frequency current correlations are measured at the output 
ports $3$ and $4$. The electron states, at energy $E$, 
are labeled by the indices 
$(j,\alpha)$ where $j=1,\cdots,4$ labels the ports of the e-HOM interferometer, 
and where $\alpha$ is the composite index $\{\ell,s\}$ with $s$ referring 
to electron spin component along the 
quantization axis, and $\ell$ referring to the orbital channel. Following the 
Landauer-B\"uttiker scattering formulation of quantum transport~\cite{buttiker} 
we introduce a set of fermionic operators for the incoming $a_{j,\alpha}(E)$ 
and outgoing $b_{j,\alpha}(E)$ states. They are connected via 
the scattering  matrix by the relation 
\begin{eqnarray}
	b_{j,\alpha}(E) = \sum_{j^\prime} \; 
	{\cal S}_{j,j^\prime}^{(\alpha)}(E) \; a_{j^{\prime},\alpha}(E)\;,
\label{Nscattering}
\end{eqnarray}
where we assume that the spin is conserved in the scattering 
process and that there is no channel mixing. The phase shifts at the input port 
2 are described by the mapping 
\be
	a_{2,\alpha}(E)\rightarrow e^{i\varphi_{\alpha}} a_{2,\alpha}(E)\;,
\label{Ntransformation}
\enn 
where $\varphi_{\alpha}$  depends on some externally controlled parameters. 
These transformations can be implemented by introducing local gate voltages 
and/or magnetic fields at the input port 2, allowing for an independent 
control of orbital- and spin-related phases, respectively.
In particular, by introducing local electric gatings $V_\ell$ plus an effective magnetic 
anisotropy $B_{\rm eff}$ (due to, e.g., spin-orbit coupling~\cite{egues02-05}), 
for small perturbations we find $\varphi_{\alpha} = 
(eV_\ell/E_\ell) (k_\ell L_0/2)+s(\mu B_{\rm eff}/E_\ell) (k_\ell L_1/2)$, 
with $k_\ell$ the electronic wave-number along the orbital channel $\ell$, 
$E_\ell=(\hbar^2/2m)k_\ell^2$, $\mu$ the electronic magnetic moment, $s=\pm 1$ the spin, 
and $L_0$ ($L_1$) the length on which $V_\ell$ ($B_{\rm eff}$) acts.
For simplicity we will assume a symmetric BS which does not suffer from backscattering. 
Hence the scattering matrix ${\cal S}$  has a block structure of the form 
\be
	{\cal S} = \left( \begin{array}{cc} 0 & \openone \\
	\openone & 0 \end{array} \right) 
	\otimes \hat{s}^{(\alpha)}(E),
\nonumber
\enn
where 
$ \hat{s}^{(\alpha)}(E)$ describes the transformation of
$a_{1,\alpha}(E), a_{2,\alpha}(E)$ into $b_{3,\alpha}(E), b_{4,\alpha}(E)$
and it is expressed by the matrix
\be
	\hat{s}^{(\alpha)}(E) \equiv 
	\left(\begin{array}{cc}
	\sqrt{1-T} &      \sqrt{T} \; e^{i \varphi_{\alpha}} \\
	\sqrt{T}   &   - \sqrt{1-T} \; e^{i \varphi_{\alpha}} \\
	\end{array} \right)
	\;,
\label{matrix}
\enn
where $T$ is the BS transmissivity that could be controlled by, e.g., a gate
voltage.\cite{LOYT98}

The current operator of the $j$-th port is defined as~\cite{lesovik}
\be
	I_j(t) \equiv \frac{e}{h\nu} \sum_{E,\omega, \alpha} e^{-i\omega t} 
 	[ b^\dag_{j,\alpha}(E)b_{j,\alpha}(E+\hbar\omega) \nonumber \\ 
	- a^\dag_{j,\alpha}(E)a_{j,\alpha}(E+\hbar\omega) ] \;,
\label{current}
\enn
where $\nu$ is the  density of states of the leads and where
a discrete spectrum has been considered to ensure  a proper regularization of 
the current correlations.
The zero-frequency current correlations are defined as
\begin{equation}
	S_{jj'} = \lim_{{\cal T} \to \infty} \frac{h \nu }
{{\cal T}^2}\int_{0}^{\cal T} dt_1 \int_{0}^{\cal T} dt_2 \; 
	\langle  \delta I_{j}(t_1)  \delta I_{j'}(t_2) \rangle ,  
\label{currentcurrent} 
\end{equation}
where the average $\langle \cdots \rangle$ 
is taken over the incoming electronic state, ${\cal T}$ 
is the measurement 
time and $\delta I_{j} = I_{j} - \langle I_{j}\rangle$. 
We now show that by studying the 
functional dependence of Eq.~(\ref{currentcurrent})
upon the  phase shifts one can detect the presence
of entanglement in the input state of the interferometer.

\section{Entanglement detection}
\label{sec:witness}

\begin{figure}[t]
\begin{center}
\includegraphics[scale=0.42]{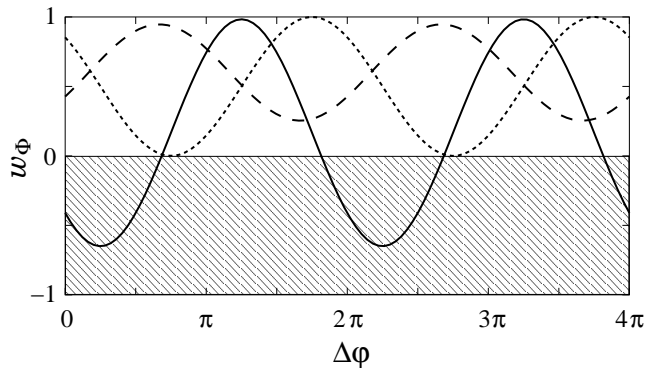}
\end{center}
\caption{
Entanglement detection scheme applied to the 
	case of two-level electronic states with component
	$\alpha=\uparrow,\downarrow$. 
	The quantity $w_{\Phi}$ is plotted as a function of 
	$\Delta\varphi=\varphi_{\uparrow}-\varphi_{\downarrow}$.
	A negative $w_{\Phi}$ (shaded area) signals the 
	entanglement in the input state (\ref{statoin}). Results are shown
	for: (solid line) a detectable entangled state with 
	$\Phi_{\uparrow,\downarrow}=1/\sqrt{2}$, 
	$\Phi_{\downarrow,\uparrow}=e^{i 3\pi/4}/\sqrt{3}$, 
	$\Phi_{\uparrow,\uparrow}=\Phi_{\downarrow,\downarrow}=
	1/\sqrt{12}$; 
	(dashed line) an undetectable entangled state  with 
	$\Phi_{\uparrow,\downarrow}=\Phi_{\uparrow,\uparrow}=
	\Phi_{\downarrow,\downarrow}=\sqrt{3/10}$, 
	$\Phi_{\downarrow,\uparrow}=e^{-i2\pi/3 }/\sqrt{10}$; 
	and (dotted line) a separable state with
	$\Phi_{\uparrow,\uparrow}=\Phi_{\uparrow,\downarrow}=1/2$, and
	$\Phi_{\downarrow,\downarrow}= \Phi_{\downarrow,\uparrow} =e^{i \pi/4}/2$.
}
\label{f:figu1}
\end{figure}

Consider the {\em two-electron} case in which, for a given energy $E$,
one electron per port enters the interferometer from $1$ and $2$. The 
most general \emph{pure} input state of this form can be expressed as
\be
	|\Psi\rangle = \prod_{E}\sum_{\alpha,\beta} \Phi_{\alpha,\beta} \; \; 
	a_{1,\alpha}^\dag(E) a_{2,\beta}^\dag(E) \; |0\rangle 
\label{statoin}\;,
\enn
where the product is taken for energies in the range $0<E<eV$, as though 
ports 1 and 2 were kept at a voltage bias $V$ with respect to ports 3 and 4.
Furthermore, $|0\rangle$ is the Fermi sea at zero temperature and $\Phi_{\alpha,\beta}$
is the two-electron amplitude which we assume to be independent
of $E$ and  satisfying the  normalization condition
$\sum_{\alpha,\beta} |\Phi_{\alpha,\beta}|^2 =1$.
One can easily verify that, independently of $\Phi_{\alpha,\beta}$, the average current  is constant and equal 
to $e^2V/h$.  A straightforward calculation of the Fano factor 
$F_{jj^\prime}=S_{j j^\prime}/(2e\sqrt{\langle I_j\rangle
\langle I_{j^\prime} \rangle} )$  leads to the expression
\be
	F_{34} =
	- T (1-T) \; (1 - w_{\Phi})
\label{cc}\;,
\enn
where
\be
	w_{\Phi}
	\equiv \;  \sum_{\alpha,\beta} 
	[\Phi_{\alpha,\beta}]^* \; 
	\Phi_{\beta,\alpha} \; 
	e^{i ( \varphi_{\alpha} - \varphi_{\beta})}\; 
\label{theg}
\enn
is a (real) quantity  which depends on the controllable 
set of phases $\{ \varphi_\alpha\}$. 
By using the Chauchy-Schwartz
 inequality and the normalization condition of $\Phi_{\alpha,\beta}$  it follows 
that for generic input state~(\ref{statoin}) one has
$-1 \leqslant  w_\Phi \leqslant 1 $.
However, if $|\Psi \rangle$ in Eq.~(\ref{statoin}) 
is a separable state with respect to  
the input ports $1$ and $2$, it is possible to show
that $w_\Phi$ is non negative. In fact, for a separable
state~(\ref{statoin}), the two-electron amplitude factorizes as
$
	\Phi_{\alpha,\beta,\rm sep} = \chi^{(1)}_{\alpha} \chi^{(2)}_{\beta} 
$
with $\chi^{(1)}_{\alpha}$ ($\chi^{(2)}_{\beta}$) being the amplitude associated 
to the incoming electron of port 1 (2). Replacing this expression 
in Eq.~(\ref{theg}) we get
\be
	w_{\Phi, \rm sep}\equiv \left| 
	\sum_{\alpha} \left[\chi^{(1)}_{\alpha} \right]^* \chi^{(2)}_\alpha 
	\; e^{i\varphi_{\alpha}}\right|^2 \geqslant 0 \;.
\label{thegsep}
\enn
By convexity the same result applies also to any separable mixed state 
of the form $\rho_{\rm sep} = \sum_i \; p_i \; |\Psi_{\rm sep}(i)
\rangle \langle \Psi_{\rm sep}(i)|$, with $p_i \ge 0$. Negative values 
of $w_{\Phi}$ are hence a direct evidence of the presence of entanglement in the input 
state, i.e.\cite{NEWNOTA}
\begin{equation}
 	{\rm If} \;\;\;   
	F_{34}< -T(1-T)
	\;\;  
	\Rightarrow \;\; {\rm The} 
	\;  {\rm state} 
	\;  {\rm is }
	\;  {\rm entangled.}
\label{wit}
\end{equation}
Notice that the inverse implication does not hold: indeed there exist
entangled states~(\ref{statoin}) which have $w_{\Phi}$ positive and hence
$F_{34}\geqslant -T(1-T)$ (see Fig.~\ref{f:figu1} for an example).
The quantity $w_{\Phi}$ therefore acts as an 
{\em entanglement witness}~\cite{horodecki}
for the class of two-electron states analyzed here.
The condition of Eq.~(\ref{wit}) 
has been derived in Ref.~\onlinecite{burkardloss} 
for the spin entangled case. Here we extended its 
validity to a generic multi-mode 
entangled input state as defined in Eq.~(\ref{statoin}).
Equation~(\ref{wit}) 
shows that a simple measurement of the Fano factor may be sufficient to 
ascertain if a given state is  entangled. Once the transmissivity $T$ 
of the BS is known and the phases $\{\varphi_\alpha\}$ are tunable, the test~(\ref{wit}) 
is within current available experimental abilities.

As an example we illustrate in Fig.~\ref{f:figu1} the detection 
of the entanglement between two-level electronic states with (pseudo)spin 
$\alpha=\uparrow,\downarrow$ at, e.g., the output of one of the spin or orbital 
entanglers so far proposed.\cite{reviews} 
By defining $\Phi_{\alpha,\beta} \equiv |\Phi_{\alpha,\beta}| e^{i \eta_{\alpha \beta}}$
we rewrite the quantity in Eq.~(\ref{theg}) as 
\begin{eqnarray}
	w_\Phi&=& |\Phi_{\uparrow,\uparrow}|^2+ |\Phi_{\downarrow,\downarrow}|^2  
	\label{Nexample}\\
	&+& 2 |\Phi_{\uparrow,\downarrow} \Phi_{\downarrow,\uparrow}| 
	\cos(\Delta \varphi + \Delta \eta) \nonumber \;,
\end{eqnarray}
with 
$\Delta \eta \equiv \eta_{\downarrow \uparrow}- \eta_{\uparrow \downarrow}$ and 
$\Delta \varphi \equiv \varphi_\uparrow-\varphi_\downarrow$. Entanglement 
can be guaranteed only for those input states leading to a negative $w_\Phi$ 
(shaded area in Fig.~\ref{f:figu1}) for some value of $\Delta \varphi$.
We show results for three typical states that can enter the BS analyzer: a detectable 
entangled state (solid line), an undetectable entangled state (dashed line), and a 
separable state (dotted line). 

\section{Lower Bound on the entanglement of formation}~\label{sec:lb}

Our  analysis of the generic input state defined in Eq.(\ref{statoin}) proceeds further.
By generalizing an argument presented in  Ref.~\onlinecite{burkardloss}, 
in this section  we show that the current noise measurement of Eq.~(\ref{cc}) is
related to the entanglement of formation~\cite{BENNETT} of the input state of the 
e-HOM interferometer.  Fundamental ingredients of our analysis are the 
 $d$-dimensional {\em generalized Werner states}~\cite{WERNER,WERNER0} 
whose properties  are briefly reviewed in the following for completeness.

\subsection{Generalized Werner stated and generalized twirling transformations}
\label{sec:genwerner}
Consider a system composed of two $d$-dimensional subsystems $1$ and $2$ and 
characterized by a separable canonical basis  $|\alpha \beta \rangle \equiv 
|\alpha \rangle_1\otimes |\beta \rangle_2$ with $\alpha,\beta \in \{1, \cdots, d\}$. 
Following the notation of Ref.~\onlinecite{LEE} we introduce the  generalized Werner 
states~\cite{WERNER,WERNER0}
\be
	\sigma_W & = &\frac{2(1-W)}{d (d+1)} 
	\left( \sum_{\alpha=0}^d |\alpha \alpha \rangle \langle \alpha\alpha | 
	+ \sum_{\alpha < \beta } |\Psi^{(+)}_{\alpha \beta}
	\rangle\langle \Psi^{(+)}_{\alpha \beta}| \right)\nonumber \\
	&+&  \frac{2W}{d (d-1)} \left(\sum_{\alpha <
	\beta } |\Psi^{(-)}_{\alpha \beta}\rangle\langle \Psi^{(-)}_{\alpha \beta}| \right)\;,
\label{werner1}
\enn
with  $W\in [0,1]$ and 
\be
	|\Psi^{(\pm)}_{\alpha \beta }
	\rangle \equiv \frac{ |\alpha \beta \rangle \pm |\beta \alpha\rangle }{\sqrt{2}}\;. 
\label{Nsinglet}
\enn
For $d=2$ Eq.~(\ref{werner1}) reduces to the standard definition of a qubit Werner 
state.\cite{WERNER0} The entanglement   of formation~\cite{BENNETT} $E_f$  of the 
density matrix $\sigma_W$ has been computed in  Ref.~\onlinecite{WERNER}.
Indeed one can verify that  these states are entangled if and only if $W>1/2$ and that 
$E_f(\sigma_W) = {\cal E}(W)$ with  ${\cal E}(x)$ being the function
\begin{eqnarray}
	{\cal E}(x) \equiv \left\{  \begin{array}{lll}
	0 & & \mbox{for $x\in[0,1/2]$} \\
	H\left( \frac{1+2\sqrt{x(1-x)}}{2}\right)
	& & \mbox{for $x\in\; ]1/2,1]$} \;,
\label{EF}
\end{array}
\right.
\end{eqnarray}
(here $H(x)=-x\log_2 x - (1-x) \log_2 (1-x)$ is the binary Shannon entropy).
Another important feature of the Werner set~(\ref{werner1}) is the fact that any 
input state $\rho$ of the system can be transformed into one of the $\sigma_W$ by 
means of Local  Operations and Classical Communications (LOCC),
i.e.
\begin{eqnarray}
	\rho \longrightarrow \sigma_{W(\rho)}\;,
\label{mappa}
\end{eqnarray} 
with $W(\rho)$ being the expectation value  of the observable $\sum_{\alpha <\beta} 
|\Psi^{(-)}_{\alpha\beta}\rangle\langle \Psi^{(-)}_{\alpha\beta}|$ 
on the input state $\rho$, i.e.
\be
	W(\rho) 
	= \sum_{\alpha <\beta} \langle \Psi^{(-)}_{\alpha 
	\beta }| \rho |\Psi^{(-)}_{\alpha \beta}\rangle \label{www1}\;. 
\enn
The map~(\ref{mappa})  is a {\em generalized twirling transformation}~\cite{WERNER,LEE} 
which can be expressed as follows,
\be
	T_{\text{wer}}(\rho) \equiv \int dU \; (U\otimes U)\;\;  \rho \; \; 
	(U^\dag \otimes U^\dag)
\label{tw}
\enn
with $U$ being a generic unitary operator on the subsystem and $d U$ being a 
proper measure in the space of such unitary transformations.
It can be shown that Werner states $\sigma_W$ are  
fix-points of $T_{\text{wer}}$, i.e. $T_{\text{wer}}(\sigma_W) = \sigma_W$.

By exploiting the properties of $T_{\text{wer}}$, Eq.(\ref{mappa}) yields a lower 
bound to the entanglement of formation of the density matrix $\rho$. In fact, 
since the entanglement of a state does not increase under LOCC transformations, we 
must have 
\be
	E_f(\rho) &\geqslant& E_f(\sigma_{W(\rho)}) = 
	{\cal E}(W(\rho)) \;,
\label{ef}
\enn
with ${\cal E}(x)$ as in Eq.~(\ref{EF}). Eq.(\ref{ef}) shows that the quantity $W(\rho)$
of Eq.~(\ref{www1}) can be used to bound the entanglement of $\rho$.
In Sec.~\ref{sec:connection} we shall prove that the Fano factor~(\ref{cc})
with all $\varphi_\alpha=0$ provides a natural way of measuring the quantity~(\ref{www1}) 
for two-electron input states.

The bound~(\ref{ef}) can be strengthen by noticing that the above derivation 
holds for all choices of the separable canonical basis $|\alpha \beta\rangle$.
Indeed for any inequivalent definition of such basis one gets a new set
of Werner density matrices and a corresponding  new LOCC twirling transformation.
Therefore, since ${\cal E}(x)$ is non decreasing, Eq.~(\ref{ef}) can be replaced by
\be
	E_f(\rho) &\geqslant& 
	{\cal E}(\overline{W}(\rho)) \;,
\label{efmod}
\enn
where $\overline{W}(\rho)$ is the maximum value of the quantity in  Eq.~(\ref{www1})
with respect to any possible choice of the separable basis $|\alpha\beta\rangle$.
To analyze the performance of the e-HOM set-up we can focus only on those basis 
which are generated from $|\alpha\beta\rangle$ by applying to subsystem $2$ the 
unitary transformation 
\begin{eqnarray}
	|\beta\rangle_2 \rightarrow V_2 |\beta \rangle_2 = 
	e^{-i\varphi_\beta} 
	|\beta\rangle_2 \label{Nphaseshift} \;,
\end{eqnarray}
with $\varphi_\beta$ being one of the controllable phases of the interferometer.
The resulting modified Werner states $\tilde{\sigma}_W$
are  obtained from Eq.~(\ref{werner1}) by replacing 
$|\Psi_{\alpha\beta}^{(\pm)}\rangle$ with the vectors 
\be
	|\tilde{\Psi}^{(\pm)}_{\alpha \beta }\rangle \equiv 
	(e^{-i\varphi_\beta} 
	|\alpha \beta \rangle \pm e^{-i\varphi_\alpha}
	|\beta \alpha\rangle)/\sqrt{2} \;,
\label{NNsinglet}
\enn
while the corresponding modified twirling transformation is obtained by properly 
concatenating $T_{\text{wer}}$ of  Eq.~(\ref{tw}) with the unitary
phase shifts transformations~(\ref{Nphaseshift}), i.e.
\be
	\tilde{T}_{\text{wer}}(\rho) \equiv \int dU \; (U \otimes 
	V_2 U V_2^\dag )\;\;  \rho \; \; 
	(U^\dag \otimes V_2 U^\dag V_2^\dag ) \;.
\nonumber 
\enn
In this context, the $\overline{W}(\rho)$ of Eq.~(\ref{efmod}) is the maximum of 
\be
	\tilde{W}(\rho)
	= \sum_{\alpha <\beta} \langle \tilde{\Psi}^{(-)}_{\alpha 
	\beta }
	| \rho |\tilde{\Psi}^{(-)}_{\alpha \beta} 
	\rangle \label{www111}\;, \enn
for all choices of the phases $\{\varphi_\alpha\}$.

\subsection{Connection with the Fano factor}\label{sec:connection}

In the notation of Eq.~(\ref{statoin})  the states  $|\alpha\beta\rangle$ 
of Eq.~(\ref{werner1}) can be identified with 
\begin{eqnarray}
	|\alpha\beta\rangle = 
	\prod_E a_{1,\alpha}^\dag(E) a_{2,\beta}^\dag(E) \; |0\rangle \;,
\label{Nstatoin}
 \end{eqnarray}
(this is possible since $a_1$ and $a_2$ refer to independent annihilation 
operators). It is then easy to verify that, assuming $\rho$ 
to be the input state~(\ref{statoin}) and exploiting the
normalization condition of the two-electron amplitude $\Phi_{\alpha
\beta}$, the right hand  side of Eq.~(\ref{www111}) becomes  
\be
	\tilde{W}(\Psi)
	= \frac{ 1 - w_\Phi}{2} \;,
\label{www333}
\end{eqnarray}
with $w_\Phi$ being the quantity in Eq.~(\ref{theg}).
In particular, setting all the interferometer phases $\{\varphi_\alpha\}$ to zero,
we get $W(\rho)$ of Eq.~(\ref{www1}). Equation (\ref{www333}) has been explicitly derived
for pure input states~(\ref{statoin}). It  can however be  generalized to 
any mixture by linearity. Combining Eqs.~(\ref{www333}) and ~(\ref{cc}) one gets
\be
	\tilde{W}(\Psi)
	= - \frac{F_{34}}{2T(1-T)}\;.
\label{www555}
\end{eqnarray}
This shows that by measuring the Fano factor 
one can determine $\tilde{W}(\Psi)$ and hence derive a lower 
bound for the entanglement of formation through Eq.~(\ref{efmod}).

\section{Conclusion}
\label{sec:conclusion}

In this paper we introduced the fermionic analog of the Hong-Ou-Mandel interferometer. 
Within this context we discussed the possibility of detecting the presence of multi-mode 
entanglement for a class of two-electron inputs states by performing current noise measurements. 
Following Ref.~\onlinecite{burkardloss}  we also showed that Fano factors measurements
provide a natural way of lower bounding the  entanglement of formation of such  incoming states.

We end with a comparison of the results discussed above with the
standard optical HOM interferometer~\cite{HOM} (see also Ref.~\onlinecite{LASER}).
The optical analog of the  two-electron input state~(\ref{statoin}) is formally obtained by
replacing $a_{j,\alpha} (E)$ with bosonic annihilation operators $a_{j}(k)$ associated with
collinear electromagnetic modes of frequency $\omega_k$ and propagating along the
$j$-th optical path,\cite{NOPTICAL} i.e.
\be
	|\Psi\rangle = \sum_{k_1,k_2} \Phi_{k_1,k_2} \, 
	a_{1}^\dag(k_1)\; a_{2}^\dag(k_2) \; |0\rangle 
\label{NNNstatoin}\;,
\enn
where $|0\rangle$ is now the electromagnetic vacuum
(notice the absence of the productory with respect to $E$).
Analogously, the current noise~(\ref{currentcurrent})
is replaced by the function
\begin{equation}
	S_{34} = \lim_{{\cal T} \to \infty} \frac{1}
	{{\cal T}^2}\int_{0}^{\cal T} dt_1 \int_{0}^{\cal T} dt_2 \; 
	\langle  \delta I_{3}(t_1)  \delta I_{4}(t_2) \rangle ,  
\label{Bcurrentcurrent} 
\end{equation}
which measures coincidence counts fluctuations  of the photo-detectors located at the
output ports $3$ and $4$.\cite{notaultima}
Replacing Eq.~(\ref{NNNstatoin}) into Eq.~(\ref{Bcurrentcurrent}) 
gives
\begin{eqnarray}
	F_{34} \equiv \frac{S_{34}}{2 \sqrt{\langle I_{3} \rangle \langle I_{4}\rangle}}
	= -T(1-T) (1 + w_\Phi) \label{FANOBOSE}\;,
\end{eqnarray}
which should be compared with Eq.~(\ref{cc}) of the  electronic case (here $w_\Phi$ 
is as in Eq.~(\ref{theg})).  The sign difference in front of the function $w_\Phi$  
is a typical signature of the bunching behavior of bosonic particles.\cite{LASER}
Nevertheless also in the bosonic case one can use the negativity of $w_{\Phi}$ as 
a signature of entanglement for the incoming two-photons states~(\ref{NNNstatoin}).

\acknowledgments
This work was supported by the European Community (grants RTN-Spintronics, RTNNANO and EUROSQIP), 
by MIUR-PRIN, and by the Quantum Information research program of Centro di Ricerca 
Matematica ``Ennio De Giorgi'' of Scuola Normale Superiore.



\begin{thebibliography}{99}
\bibitem{NEWNOTA}
	Here there is a 
	subtlety.  Due to the non linear dependency upon the currents 
	$\langle I_3 \rangle$ and  $\langle I_4\rangle$ the Fano 
	factor $F_{34}$ is in general not a linear function of the
	input state. Therefore $F_{34}$  
	of a convex combination is not necessarily equal to the convex
	combination of the Fano factors 
	 of its components. In our case however
	$\langle I_3 \rangle$ and  $\langle I_4\rangle$ are constant
	and   do not 
	depend on $|\Psi\rangle$: therefore $F_{34}$ is linear on $\rho$.
\bibitem{notaultima}Here for $j=3,4$, $I_{j}(t)$ is 
	the intensity operator which counts
	the number of photons impinging on a photo-detector
	located at the $j$-th output port.  It is given by
	$I_{j}(t)=\sum_{k_1} \sum_{k_2}e^{i (\omega_{k_1}-\omega_{k_2})t} 
	\label{Nphotonumber} b_j^\dag(k_1)  b_j(k_2)$, where 
	$b_j(k)$ is the corresponding bosonic annihilation operator 
	which can be expressed in term of $a_{1,2}(k)$ through
	a scattering matrix similar to~(\ref{matrix}).
\end{thebibliography}
\end{document}